\documentclass[aps,prl, amsmath, showpacs, preprintnumbers,superscriptaddress, twocolumn, amssymb]{revtex4}

\usepackage{graphicx}
\usepackage{mathptmx, textcomp}
\usepackage[latin1]{inputenc}
\usepackage{color}
\begin{document}
\title{Evidence for universal four-body states tied to an Efimov trimer}
\author{F. Ferlaino}
\author{S. Knoop}
\author{M. Berninger}
\author{W. Harm}
\affiliation{Institut f\"ur Experimentalphysik and Zentrum f\"ur Quantenphysik, Universit\"at
 Innsbruck, 
 6020 Innsbruck, Austria}
\author{J. P. D'Incao}
\affiliation{Institut f\"ur Quantenoptik und Quanteninformation,
 \"Osterreichische Akademie der Wissenschaften, 6020 Innsbruck,
 Austria}
 \affiliation{JILA, University of Colorado and NIST, Boulder, Colorado 80309-0440, USA}
\author{H.-C. N\"{a}gerl}
\affiliation{Institut f\"ur Experimentalphysik and Zentrum f\"ur Quantenphysik, Universit\"at
 Innsbruck, 
 6020 Innsbruck, Austria}
\author{R. Grimm}
\affiliation{Institut f\"ur Experimentalphysik and Zentrum f\"ur Quantenphysik, Universit\"at
 Innsbruck, 
 6020 Innsbruck, Austria}
\affiliation{Institut f\"ur Quantenoptik und Quanteninformation,
 \"Osterreichische Akademie der Wissenschaften, 6020 Innsbruck,
 Austria}

\date{\today}

\pacs{03.75.-b, 34.50.Cx, 67.85.-d, 21.45.-v}

\begin{abstract}
We report on the measurement of four-body recombination rate coefficients in an atomic gas.
Our results obtained with an ultracold sample of cesium atoms at negative scattering lengths show a resonant enhancement
of losses and provide strong evidence for the existence of a pair of four-body states, which is strictly connected to Efimov trimers via  universal relations. Our findings confirm recent theoretical predictions and demonstrate the enrichment  of the Efimov scenario when a fourth particle is added to the generic three-body problem.
\end{abstract}

\maketitle


Few-body physics produces bizarre and counterintuitive phenomena, with the Efimov effect representing the major paradigm of the field \cite{Efimov1970ela}.
Early in the 1970s, Efimov found a solution to the quantum three-body problem, predicting the existence of an infinite series of universal weakly bound three-body states. Surprisingly, these Efimov trimers can even exist under conditions where a weakly bound dimer state is absent \cite{Jensen2004sar, Kohler2006poc, Braaten2006uif}. An essential prerequisite for the Efimov effect is a large two-body scattering length $a$, far exceeding the characteristic range of the interaction potential. Ultracold atomic systems with tunable interactions \cite{Chin2008fri} have opened up unprecedented possibilities to explore such few-body quantum systems under well controllable experimental conditions. In particular, $a$ can be made much larger than the van der Waals length $r_{\rm vdW}$ \cite{vdW}, the range of the interatomic interaction.

In the last few years, signatures of Efimov states have been observed in ultracold atomic and molecular gases of cesium atoms \cite{Kraemer2006efe,Knoop2008ooa}, and recently in three-component Fermi gases of $^6$Li \cite{ottenstein2008cso,Huckans2008}, in a Bose gas of $^{39}$K atoms \cite{zaccantiDAMOP}, and in mixtures of $^{41}$K and $^{87}$Rb atoms \cite{barontini2009ooh}. In all these experiments, Efimov states manifest themselves as resonantly enhanced losses, either in atomic three-body recombination or in atom-dimer relaxation processes. The recent observations highlight the universal character of Efimov states, and they also point to a rich playground for future experiments.

\begin{figure}
 \includegraphics[width=8.0cm] {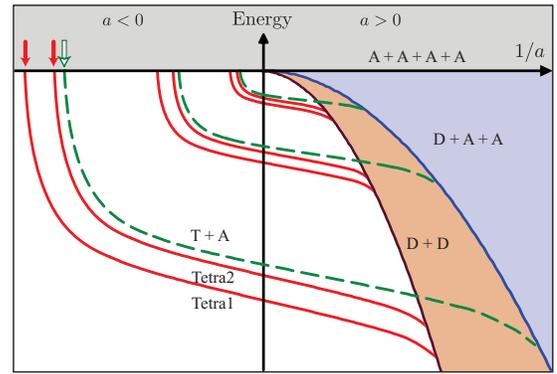}
 \vspace{-7mm}
 \caption{(color online) Extended Efimov scenario describing a universal system of four identical bosons; Energies are plotted as a function of the inverse scattering length. The red solid lines illustrate the pairs of universal tetramer states (Tetra1 and Tetra2) associated with each Efimov trimer (T).
 For illustrative purposes, we have artificially reduced the universal Efimov scaling factor from 22.7 to about 2. The shaded regions indicate the scattering continuum associated with the relevant dissociation threshold. The four-body threshold is at zero energy and refers to four free atoms (A+A+A+A). In the $a>0$ region, the dimer-atom-atom threshold (D+A+A) and the dimer-dimer threshold (D+D) are also depicted. The weakly bound dimer, only existing for $a\gg r_{\rm vdW}>0$, has universal halo character and its binding energy is given by $\hbar^2/(ma^2)$ \cite{Jensen2004sar, Ferlaino2008cbt}. The open arrow marks the intersection of the first Efimov trimer (T) with the atomic threshold, while the filled arrows indicate the corresponding locations of the two universal tetramer states.}
 \label{fig1}
\end{figure}

As a next step in complexity, a system of four identical bosons with resonant two-body interaction challenges our understanding of few-body physics.
The extension of universality to four-body systems has been attracting increasing interest both in theory  \cite{platter2004fbs, Yamashita2006fbs, hanna2006eas, Hammer2007upo, Wang2008etf, vonStecher2008fbl} and experiment \cite{Ferlaino2008cbt}. A particular question
under debate is the possible relation between universal three- and four-body states \cite{platter2004fbs, Yamashita2006fbs, hanna2006eas, Hammer2007upo, vonStecher2008fbl}. In this context, Hammer and Platter predicted the four-body system to support universal tetramer states in close connection with Efimov trimers \cite{Hammer2007upo}.

Recently, von Stecher, D'Incao, and Greene presented key predictions for universal four-body states \cite{vonStecher2008fbl}. For each Efimov trimer, they demonstrate the existence of a pair of universal tetramer states according to the conjecture of Ref.\,\cite{Hammer2007upo}. Such tetramer states are tied to the corresponding trimer through simple universal relations that do not invoke any four-body parameter \cite{platter2004fbs, hanna2006eas, vonStecher2008fbl}. The authors of Ref.\,\cite{vonStecher2008fbl} suggest resonantly enhanced four-body recombination in an atomic gas as a probe for such universal tetramer states. They also find hints on the existence of one of the predicted four-body resonances by reinterpreting our earlier recombination measurements on $^{133}$Cs atoms at large negative scattering lengths \cite{Kraemer2006efe}. In this Letter, we present new measurements on the Cs system dedicated to four-body recombination in the particular region of interest near a triatomic Efimov resonance. Our results clearly verify the central predictions of Ref.\,\cite{vonStecher2008fbl}. We observe two loss resonances as a signature of the predicted tetramer pair and we find strong evidence for the four-body nature of the underlying recombination process.

\begin{figure}
 \includegraphics[width=8.5cm] {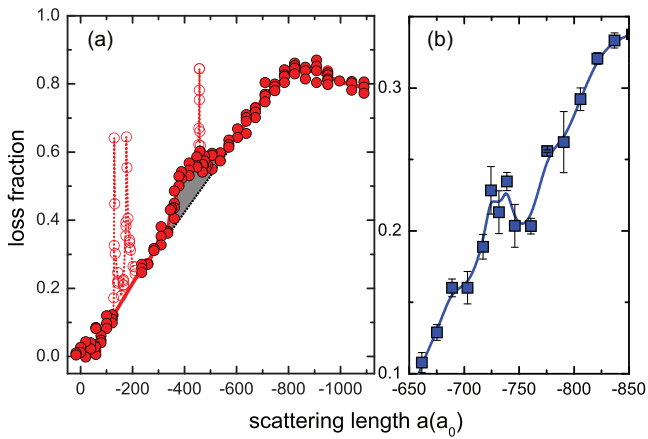}
 \caption{(color online) Recombination losses in an ultracold sample of Cs atoms. (a) Loss fraction for a 50-nK sample after a storage time of 250~ms. Here we present all individual measurements to give an impression of the scatter of our data.The broad maximum at about $-870 a_0$ is caused by a triatomic Efimov resonance \cite{Kraemer2006efe} and the shaded area highlights the resonant loss enhancement that we attribute to the four-body state Tetra1. The three very narrow loss features (open circles) are caused by known $g$-wave Feshbach resonances \cite{Chin2004pfs}, which are irrelevant in the present context. (b) Loss fraction for a 30-nK sample after a storage time of 8~ms. Each data point represents  the
average values resulting from five individual measurements for a given $a$ together with their statistical errors. The loss enhancement at around $-730 a_0$ is caused by the state Tetra2. The solid lines are spline interpolations guiding the eye.}
 \label{fig2}
\end{figure}
The four-body extended Efimov scenario \cite{Hammer2007upo,vonStecher2008fbl} is schematically  illustrated in Fig.\,\ref{fig1}, where the tetramer states (Tetra1 and Tetra2) and the relevant thresholds are depicted as a function of the inverse scattering length $1/a$.
Within the four-body scenario, the Efimov trimers (T) are associated with trimer-atom thresholds (T+A, dashed lines). The pair of universal tetramer states (solid lines) lies below the corresponding T+A threshold. The four-body breakup threshold (A+A+A+A) defines zero energy and refers to the continuum of four free atoms. For completeness, we also show the $a>0$ region. Here, the picture is even richer because of the presence of the weakly bound dimer state, whichs leads to the dimer-atom-atom threshold (D+A+A) and the dimer-dimer threshold (D+D). In the four-body scenario, the tetramer states emerge at the atomic threshold for $a<0$ and connect to the D+D threshold for $a>0$.

The Efimov trimer intersects the atomic threshold at $a=a^*_{\rm T}$, which leads to the observed triatomic resonance \cite{Kraemer2006efe}. The corresponding tetramer states are predicted \cite{vonStecher2008fbl} to intersect the atomic threshold at scattering length values
\begin{equation}
\label{EqUR} a^*_{{\rm Tetra1}}\approx 0.43\ a^*_{\rm T} \hspace{2mm}
\mathrm{and}\hspace{2mm} a^*_{{\rm Tetra2}}\approx 0.9\ a^*_{{\rm T}}.
\end{equation}
These universal relations, linking three- and four-body resonances, express the fact that no additional parameter, namely the so-called four-body parameter, is needed to describe the system behavior. In contrast to the connection between universal two- and three-body systems, where a three-body parameter is required to locate the trimer states, the universal properties of the four-body system are thus directly related to the corresponding three-body subsystem.

In analogy to the well-established fact that Efimov trimers lead to loss resonances in an atomic gas \cite{Esry1999rot, Kraemer2006efe}, universal four-body states can also be expected to manifest themselves in a resonant increase of atomic losses \cite{vonStecher2008fbl}. Resonant coupling between four colliding atoms and a tetramer state ($a\simeq a^*_{{\rm Tetra}}<0$) drastically enhances four-body recombination to lower lying channels. Possible decay channels are trimer-atom, dimer-dimer, and dimer-atom-atom channels. In each of these recombination processes, we expect all the particles to rapidly escape from the trap, as the kinetic energy gained usually exceeds the trap depth.


We prepare an ultracold optically
trapped atomic sample in the lowest hyperfine sublevel $({\rm F}=3,m_{\rm F}=3)$ \cite{HF}, as described in Ref.\,\cite{Ferlaino2008cbt}.
By varying the magnetic field between 6 and 17~G, the scattering length $a$ can be tuned from $-1100$ to 0\,$a_0$ \cite{Weber2003bec}, where $a_0$ is Bohr's radius. For presenting our experimental data in the following, we convert the applied magnetic field into $a$ using the fit formula of Ref.\,\cite{Kraemer2006efe}.
After several cooling and trapping stages \cite{Weber2003bec}, the atoms are loaded into an optical trap, formed by crossing two  infrared laser beams \cite{Ferlaino2008cbt}. The trap frequencies in the three spatial directions are about $(\omega_x,\omega_y,\omega_z)=2\pi\times(10,46,65)$~Hz.  Similar to \cite{Weber2003bec}, we
support the optical trap by employing a magnetic levitation field acting
against gravity. Evaporative cooling in the levitated trap is stopped just before the onset of Bose-Einstein condensation in order to avoid implosion of the gas. For our typical temperature of 50~nK, we obtain about $8\times10^4$ non-condensed atoms with a peak density of about $7\times 10^{12}$~cm$^{-3}$.

In a first set of experiments, we record the atom number after a fixed storage time in the optical trap for variable scattering length in the $a<0$ region. Figure \ref{fig2} shows the observed losses, containing both three- and four-body contributions.
The three-body part consists of a background that follows a general $a^4$-scaling behavior \cite{Fedichev1996tbr, Esry1999rot, Weber2003tbr} and resonant losses caused by the triatomic Efimov resonance, which for a 50-nK sample was observed to occur at $a^*_{\rm T}=-870(10) a_0$ \cite{Kraemer2006efe, Naegerl2006eef}; this is consistent with the large losses shown in Fig.\,\ref{fig2}(a). Beside this expected behavior of the three-body subsystem, we clearly observe two additional loss features, one located at about $-410 a_0 $ [Fig.\,\ref{fig2}(a)] and one at about $-730 a_0$ [Fig.\,\ref{fig2}(b)].
The observation of the resonance at $-730 a_0$  is particularly demanding and requires a careful choice of parameters as the signal needs to be discriminated against the very strong background that is caused by three-body losses. Here we use a much shorter hold time of 8~ms, which is the shortest possible time required to ensure precise magnetic field control in our apparatus.

We interpret the two observed resonant loss features as the predicted pair of four-body resonances \cite{vonStecher2008fbl}.
For the resonance positions we find $a^*_{\rm Tetra1}/a^*_{\rm T} \simeq 0.47$ and $a^*_{\rm Tetra2}/a^*_{\rm T} \simeq 0.84$, which are remarkably close to the predictions of Eq.\,(\ref{EqUR}).

\begin{figure}
 \includegraphics[width=8.5cm] {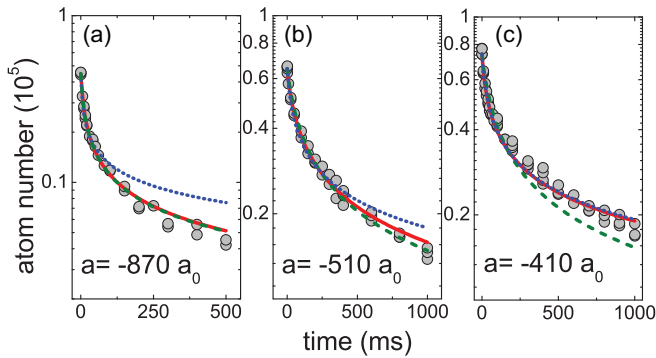}
 \caption{(color online) Time evolution of the number of atoms in an optically trapped sample. The solid lines are the fit to the data based on the full numerical solution of Eq.\,(\ref{RE}). The dashed and dotted lines correspond to a pure three-body and pure four-body decay, respectively (see text). (a) For dominant three-body collisions ($a=-870 a_0$), (b) an intermediate situation ($a=-510 a_0$), and (c) dominant four-body collisions ($a=-410 a_0$). }
 \label{fig3}
\end{figure}


In a second set of experiments, we study the time-dependence of the atomic decay in the optical trap. Here we focus on the region around the resonance at $a^*_{\rm Tetra1}\simeq -410 a_0 $, where the three-body losses are comparatively weak and thus allow for a detailed analysis of the loss curves. Representative loss measurements for three different values of $a$ are shown in Fig.\,\ref{fig3}.

The observed decay can be fully attributed to three-body and four-body  recombination collisions. This is due to the fact that inelastic two-body collisions of atoms in the lowest Zeeman sub-level are energetically suppressed, and one-body losses, such as background collisions or light-induced losses, can be completely neglected under our experimental conditions.
The corresponding differential equation for the decaying atom number reads as
\begin{equation}
\label{RE}
\dot{N}/N=-L_3 \langle n^2\rangle -L_4  \langle n^3\rangle,
\end{equation}
where $L_3$ and $L_4$ denote the three- and the four-body recombination rate coefficient, respectively. The average density is calculated by integrating the density over the volume $\langle n^2\rangle=(1/N)\int n^3 d^3\mathbf{r}$ and $\langle n^3\rangle=(1/N)\int n^4 d^3\mathbf{r}$. By considering a thermal density distribution of gaussian shape in the three-dimensional harmonic trap, we obtain $\langle n^2\rangle= n^2_p/\sqrt{27}$ and $\langle n^3\rangle= n^3_p/8$, with $n_p =\sqrt{8}N [m \overline{\omega}^2/(4\pi k_{\rm B} T)]^{3/2}$ the peak density. Here, $m$ is the atomic mass, $T$ the temperature, and $\overline{\omega}=(\omega_x\omega_y\omega_z)^{1/3}$ the mean trap frequency.
We determine the trap frequencies and the temperature by sloshing mode and time-of-flight measurements, respectively.

In general Eq.\,(\ref{RE}) is not analytically solvable.
An analytic solution can be found in the limit of either pure three-body losses or pure four-body losses.

Therefore we fit our decay curves with a numerical solution of Eq.\,(\ref{RE}), keeping both $L_3$ and $L_4$ as free parameters. Note that we have not included anti-evaporation heating \cite{Weber2003tbr} in our model because we do not observe the corresponding temperature increase in our experiments. We believe that, for the fast decay observed here, the sample may not have enough time to thermalize. 

Our experimental data clearly reveal a qualitative change of the decay curves when $a$ is tuned between  $a^*_{\rm T}$ and $a^*_{\rm Tetra1}$. Fig.\,\ref{fig3}(a) shows that for $a\approx a^*_{\rm T}$ the loss is dominated by three-body recombination; here the full numerical fitting curve follows the pure three-body solution. A different situation is found at $-410 a_0$; see Fig.\ \ref{fig3}(c). Here a pure three-body analysis cannot properly describe the observed behavior and the full numerical solution reveals a predominant four-body character. In intermediate situations, for which an example is shown in Fig.\ \ref{fig3}(b), both three- and four-body processes significantly contribute to the observed decay.

\begin{figure}
 \includegraphics[width=8.5cm] {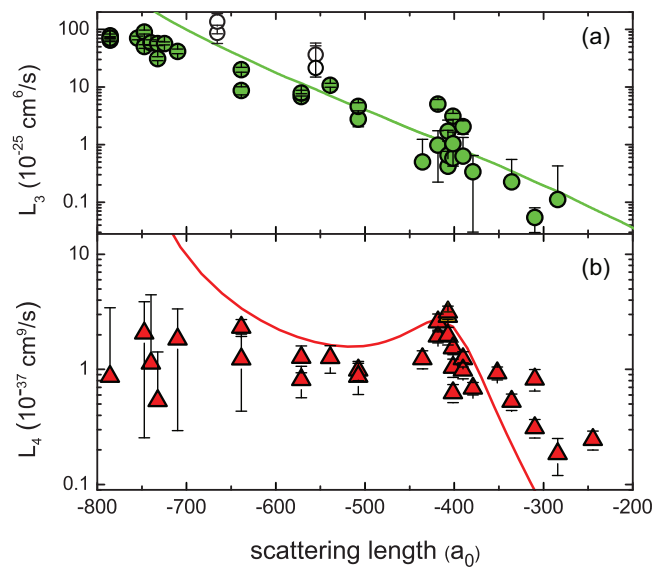}
 \caption{(color online) Loss rate coefficient for (a) three- and  (b) four-body recombination as a function of the scattering length $a$. The measurements are taken at temperatures of about 40 nK. The values are obtained by fitting the numerical solution of Eq.\,(\ref{RE}) to the decay curve. The error bars on $L_3$ and $L_4$ are the statistical uncertainties from the fit evaluated with a resampling method \cite{Westfallbook}.
 The open circles in (a) refer  to previous data at 250~nK \cite{Kraemer2006efe}. The solid curves result from the theoretical model of Refs.\,\cite{vonStecher2008fbl,Mehta}.}
 \label{fig4}
\end{figure}


From the decay curves taken at different values of $a$ we determine $L_3$ and $L_4$; the results are shown in Fig.\,\ref{fig4}(a) and (b), respectively. The three-body contribution $L_3$ follows previously observed behavior \cite{Kraemer2006efe}, as dictated by the $a^4$-scaling in combination with the Efimov effect.

Our major result is shown in Fig.\,\ref{fig4}(b), where we plot the rate coefficient $L_4$. Our data provide the first available quantitative information on $L_4$, establishing the role of four-body collisions in ultracold gases.
For $|a|<|a^*_{{\rm Tetra1}}|$, where no universal tetramer states exist, the four-body losses are typically very weak. Here, we measure $L_4\simeq 0.2 \times 10^{-37} {\rm cm}^9/$s.
With increasing $|a|$, the system undergoes a significant change in its behavior, with four-body collisions dominating the atomic decay; see Fig.\,\ref{fig3}(c).
We observe a sharp increase of $L_4$, which reaches its maximum value at $a= -412(2) a_0$.
This observation is another strong piece of evidence for the predicted universal four-body state at $a^*_{{\rm Tetra1}}$ \cite{vonStecher2008fbl}. To directly estimate the relative contributions of three- and four-body recombination, one can compare $L_3$ with $n_{0}\,L_4$, where $n_{0}\simeq 1.0 \times 10^{13}$~cm$^{-3}$ is the initial peak density at 40 nK. At resonance, $n_{0} L_4$ exceeds $L_3$ by more than one order of magnitude, with $L_3=0.7 \times 10^{-25} {\rm cm}^6/{\rm s}$ and $n_{0}L_4= 3 \times 10^{-24} {\rm cm}^6/$s.
With further increasing $|a|$, $L_4$ decreases and $L_3$ increases such that $L_3>n_{0}L_4$. For $|a|> 700 a_0$, the very fast three-body decay renders the analysis of the loss curves in terms of $L_4$ unreliable. In addition to the large statistical fit errors seen in our data in this region, other systematic error sources like a non-thermal evolution of the atomic density distribution can have a strong influence.

Figure \ref{fig4} also includes the theoretical predictions for $L_3$ and $L_4$ at 40~nK and demonstrates a remarkable qualitative agreement with our experimental results.
The theoretical approach utilizes a solution of the four-body problem in the hyperspherical adiabatic representation \cite{vonStecher2008fbl}; the derivation and associated calculations of $L_4$, adapted from \cite{Mehta}, provide the first quantitative description of the four-body recombination rate.  The calculations only require to fix the position of the triatomic Efimov resonance as determined in the previous experiment at 10~nK of Ref.\,\cite{Kraemer2006efe}.
The difference in the width and the amplitude of the four-body resonance between experimental and theoretical data may be explained by different coupling to possible decay channels.

Our work leads to important conclusions related to the concept of universality with increasing complexity. The observation of the two four-body resonances close to the predicted positions \cite{vonStecher2008fbl} points to the universal character of the underlying states. This also supports the view or Refs.\,\cite{platter2004fbs, hanna2006eas, vonStecher2008fbl}  that a four-body parameter is not required to describe the system. Universal four-body states then emerge as a genuine consequence of the Efimov spectrum. This also provides a novel way to test Efimov physics. The Efimovian character of a three-body resonance can be probed by observing the universal tetramer resonances tied to it, without the necessity to explore the full geometric scaling of Efimov physics by changing the scattering length by orders of magnitude.

While our present work has focussed on four-body phenomena at negative scattering length, a further exciting step will be the exploration of the entire four-body spectrum. For positive scattering lengths, the spectrum becomes richer and new phenomena can be expected such as resonant interactions between four-body states and two-dimer states. In this way, experiments on few-body phenomena in ultracold atoms will keep on challenging our understanding of the universal physics of a few resonantly interacting particles.

We are aware of related results in $^{39}$K, in which enhanced losses near a Feshbach resonance may be interpreted as a four-body resonance \cite{zaccanti2009}.

We acknowledge C.\,Greene and J.\,von Stecher for stimulating discussions and for sharing their theoretical curves (Fig.\,\ref{fig4}) with us.  We acknowledge support by the Austrian Science Fund (FWF) within SFB 15 (project part 16). F.\,F.\,is supported within the Lise Meitner program of
the FWF. JPD's contribution was supported in part by the NSF.



\begin{thebibliography}{29}
\expandafter\ifx\csname natexlab\endcsname\relax\def\natexlab#1{#1}\fi
\expandafter\ifx\csname bibnamefont\endcsname\relax
  \def\bibnamefont#1{#1}\fi
\expandafter\ifx\csname bibfnamefont\endcsname\relax
  \def\bibfnamefont#1{#1}\fi
\expandafter\ifx\csname citenamefont\endcsname\relax
  \def\citenamefont#1{#1}\fi
\expandafter\ifx\csname url\endcsname\relax
  \def\url#1{\texttt{#1}}\fi
\expandafter\ifx\csname urlprefix\endcsname\relax\def\urlprefix{URL }\fi
\providecommand{\bibinfo}[2]{#2}
\providecommand{\eprint}[2][]{\url{#2}}

\bibitem[{\citenamefont{Efimov}(1970)}]{Efimov1970ela}
\bibinfo{author}{\bibfnamefont{V.}~\bibnamefont{Efimov}},
  \bibinfo{journal}{Phys. Lett. B} \textbf{\bibinfo{volume}{33}},
  \bibinfo{pages}{563} (\bibinfo{year}{1970}).

\bibitem[{\citenamefont{Jensen et~al.}(2004)\citenamefont{Jensen, Riisager,
  Fedorov, and Garrido}}]{Jensen2004sar}
\bibinfo{author}{\bibfnamefont{A.~S.} \bibnamefont{Jensen}},
  \bibinfo{author}{\bibfnamefont{K.}~\bibnamefont{Riisager}},
  \bibinfo{author}{\bibfnamefont{D.~V.} \bibnamefont{Fedorov}},
  \bibnamefont{and} \bibinfo{author}{\bibfnamefont{E.}~\bibnamefont{Garrido}},
  \bibinfo{journal}{Rev. Mod. Phys.} \textbf{\bibinfo{volume}{76}},
  \bibinfo{pages}{215} (\bibinfo{year}{2004}).

\bibitem[{\citenamefont{K\"ohler et~al.}(2006)\citenamefont{K\"ohler, G\'oral,
  and Julienne}}]{Kohler2006poc}
\bibinfo{author}{\bibfnamefont{T.}~\bibnamefont{K\"ohler}},
  \bibinfo{author}{\bibfnamefont{K.}~\bibnamefont{G\'oral}}, \bibnamefont{and}
  \bibinfo{author}{\bibfnamefont{P.~S.} \bibnamefont{Julienne}},
  \bibinfo{journal}{Rev. Mod. Phys.} \textbf{\bibinfo{volume}{78}},
  \bibinfo{eid}{1311} (\bibinfo{year}{2006}).

\bibitem[{\citenamefont{Braaten and Hammer}(2006)}]{Braaten2006uif}
\bibinfo{author}{\bibfnamefont{E.}~\bibnamefont{Braaten}} \bibnamefont{and}
  \bibinfo{author}{\bibfnamefont{H.-W.} \bibnamefont{Hammer}},
  \bibinfo{journal}{Phys. Rep.} \textbf{\bibinfo{volume}{428}},
  \bibinfo{pages}{259} (\bibinfo{year}{2006}).

\bibitem[{\citenamefont{Chin et~al.}(2008)\citenamefont{Chin, Grimm, Julienne,
  and Tiesinga}}]{Chin2008fri}
\bibinfo{author}{\bibfnamefont{C.}~\bibnamefont{Chin}},
  \bibinfo{author}{\bibfnamefont{R.}~\bibnamefont{Grimm}},
  \bibinfo{author}{\bibfnamefont{P.}~\bibnamefont{Julienne}}, \bibnamefont{and}
  \bibinfo{author}{\bibfnamefont{E.}~\bibnamefont{Tiesinga}},
  \bibinfo{journal}{submitted to Rev. Mod. Phys.}  (\bibinfo{year}{2008}).

\bibitem[{vdW()}]{vdW}
\bibinfo{note}{The van der Waals length is defined as
  $r_{\rm{vdW}}=\frac{1}{2}(m C_6/\hbar^2)^{1/4}$, where $C_6$ is the van der
  Waals dispersion coefficient \cite{Kohler2006poc}. For Cs, $r_{\rm{vdW}}=100
  a_0$}.

\bibitem[{\citenamefont{Kraemer et~al.}(2006)\citenamefont{Kraemer, Mark,
  Waldburger, Danzl, Chin, Engeser, Lange, Pilch, Jaakkola, N\"agerl, Grimm
  }}]{Kraemer2006efe}
\bibinfo{author}{\bibfnamefont{T.}~\bibnamefont{Kraemer}} {\it et al.},
  \bibinfo{journal}{Nature}
  \textbf{\bibinfo{volume}{440}}, \bibinfo{pages}{315} (\bibinfo{year}{2006}).

\bibitem[{\citenamefont{Knoop et~al.}(2009)\citenamefont{Knoop, Ferlaino, Mark,
  Berninger, Sch\"{o}bel, N\"{a}gerl, and Grimm}}]{Knoop2008ooa}
\bibinfo{author}{\bibfnamefont{S.}~\bibnamefont{Knoop}} {\it et al.},
  \bibinfo{journal}{Nat. Phys.} \textbf{\bibinfo{volume}{5}},
  \bibinfo{pages}{227} (\bibinfo{year}{2009}).

\bibitem[{\citenamefont{Ottenstein et~al.}(2008)\citenamefont{Ottenstein,
  Lompe, Kohnen, Wenz, and Jochim}}]{ottenstein2008cso}
\bibinfo{author}{\bibfnamefont{T.~B.} \bibnamefont{Ottenstein}} {\it et al.},
  \bibinfo{journal}{Phys. Rev. Lett.} \textbf{\bibinfo{volume}{101}},
  \bibinfo{pages}{203202} (\bibinfo{year}{2008}).

\bibitem[{\citenamefont{Huckans et~al.}(2008)\citenamefont{Huckans, Williams,
  Hazlett, Stites, and O'Hara}}]{Huckans2008}
\bibinfo{author}{\bibfnamefont{J.~H.} \bibnamefont{Huckans}} {\it et al.},
  \bibinfo{journal}{arXiv:0810.3288}
  (\bibinfo{year}{2008}).

\bibitem[{zac()}]{zaccantiDAMOP}
\bibinfo{note}{M. Zaccanti, G. Modugno, C. D`Errico, M. Fattori, G. Roati, and
  M. Inguscio, talk at DAMOP-2008, 27-31 May 2008, State College, Pennsylvania,
  USA.}

\bibitem[{\citenamefont{Barontini et~al.}(2009)\citenamefont{Barontini, Weber,
  Rabatti, Catani, Thalhammer, Inguscio, and Minardi}}]{barontini2009ooh}
\bibinfo{author}{\bibfnamefont{G.}~\bibnamefont{Barontini}} {\it et al.},
  \bibinfo{journal}{arXiv:0901.4584}  (\bibinfo{year}{2009}).

\bibitem[{\citenamefont{Ferlaino et~al.}(2008)\citenamefont{Ferlaino, Knoop,
  Mark, Berninger, Sch\"{o}bel, N\"{a}gerl, and Grimm}}]{Ferlaino2008cbt}
\bibinfo{author}{\bibfnamefont{F.}~\bibnamefont{Ferlaino}} {\it et al.},
  \bibinfo{journal}{Phys. Rev. Lett.} \textbf{\bibinfo{volume}{101}},
  \bibinfo{pages}{023201} (\bibinfo{year}{2008}).

\bibitem[{\citenamefont{Platter et~al.}(2004)\citenamefont{Platter, Hammer, and
  Mei{\ss}ner}}]{platter2004fbs}
\bibinfo{author}{\bibfnamefont{L.}~\bibnamefont{Platter}},
  \bibinfo{author}{\bibfnamefont{H.-W.}~\bibnamefont{Hammer}}, \bibnamefont{and}
  \bibinfo{author}{\bibfnamefont{Ulf-G.}~\bibnamefont{Mei{\ss}ner}},
  \bibinfo{journal}{Phys. Rev. A} \textbf{\bibinfo{volume}{70}},
  \bibinfo{pages}{052101} (\bibinfo{year}{2004}).

\bibitem[{\citenamefont{Yamashita et~al.}(2006)\citenamefont{Yamashita, Tomio,
  Delfino, and Frederico}}]{Yamashita2006fbs}
\bibinfo{author}{\bibfnamefont{M.}~\bibnamefont{Yamashita}},
  \bibinfo{author}{\bibfnamefont{L.}~\bibnamefont{Tomio}},
  \bibinfo{author}{\bibfnamefont{A.}~\bibnamefont{Delfino}}, \bibnamefont{and}
  \bibinfo{author}{\bibfnamefont{T.}~\bibnamefont{Frederico}},
  \bibinfo{journal}{Europhys. Lett.} \textbf{\bibinfo{volume}{75}},
  \bibinfo{pages}{555} (\bibinfo{year}{2006}).

\bibitem[{\citenamefont{Hanna and Blume}(2006)}]{hanna2006eas}
\bibinfo{author}{\bibfnamefont{G.~J.} \bibnamefont{Hanna}} \bibnamefont{and}
  \bibinfo{author}{\bibfnamefont{D.}~\bibnamefont{Blume}},
  \bibinfo{journal}{Phys. Rev. A} \textbf{\bibinfo{volume}{74}},
  \bibinfo{pages}{063604} (\bibinfo{year}{2006}).

\bibitem[{\citenamefont{Hammer and Platter}(2007)}]{Hammer2007upo}
\bibinfo{author}{\bibfnamefont{H.}~\bibnamefont{Hammer}} \bibnamefont{and}
  \bibinfo{author}{\bibfnamefont{L.}~\bibnamefont{Platter}},
  \bibinfo{journal}{Eur. Phys. J. A} \textbf{\bibinfo{volume}{32}},
  \bibinfo{pages}{113} (\bibinfo{year}{2007}).

\bibitem[{\citenamefont{Wang and Esry}(2008)}]{Wang2008etf}
\bibinfo{author}{\bibfnamefont{Y.}~\bibnamefont{Wang}} \bibnamefont{and}
  \bibinfo{author}{\bibfnamefont{B.~D.} \bibnamefont{Esry}},
  \bibinfo{journal}{arXiv:0809.3779}  (\bibinfo{year}{2008}).

\bibitem[{\citenamefont{von Stecher et~al.}(2008)\citenamefont{von Stecher,
  D'Incao, and Greene}}]{vonStecher2008fbl}
\bibinfo{author}{\bibfnamefont{J.}~\bibnamefont{von Stecher}},
  \bibinfo{author}{\bibfnamefont{J.~P.} \bibnamefont{D'Incao}},
  \bibnamefont{and} \bibinfo{author}{\bibfnamefont{C.~H.}
  \bibnamefont{Greene}}, \bibinfo{journal}{arXiv:0810.3876}
  (\bibinfo{year}{2008}).

\bibitem[{\citenamefont{Chin et~al.}(2004)\citenamefont{Chin, Vuleti\'c,
  Kerman, Chu, Tiesinga, Leo, and Williams}}]{Chin2004pfs}
\bibinfo{author}{\bibfnamefont{C.}~\bibnamefont{Chin}} {\it et al.},
  \bibinfo{journal}{Phys. Rev. A} \textbf{\bibinfo{volume}{70}},
  \bibinfo{pages}{032701} (\bibinfo{year}{2004}).

\bibitem[{\citenamefont{Esry et~al.}(1999)\citenamefont{Esry, Greene, and
  Burke}}]{Esry1999rot}
\bibinfo{author}{\bibfnamefont{B.~D.} \bibnamefont{Esry}},
  \bibinfo{author}{\bibfnamefont{C.~H.} \bibnamefont{Greene}},
  \bibnamefont{and} \bibinfo{author}{\bibfnamefont{J.~P.} \bibnamefont{Burke}},
  \bibinfo{journal}{Phys. Rev. Lett.} \textbf{\bibinfo{volume}{83}},
  \bibinfo{pages}{1751} (\bibinfo{year}{1999}).

\bibitem[{HF()}]{HF}
\bibinfo{note}{${\rm F}$ and $m_{\rm F}$ indicate the hyperfine and projection
  quantum number, respectively.}

\bibitem[{\citenamefont{Weber et~al.}(2003{\natexlab{a}})\citenamefont{Weber,
  Herbig, Mark, N\"agerl, and Grimm}}]{Weber2003bec}
\bibinfo{author}{\bibfnamefont{T.}~\bibnamefont{Weber}} {\it et al.},
  \bibinfo{journal}{Science} \textbf{\bibinfo{volume}{299}},
  \bibinfo{pages}{232} (\bibinfo{year}{2003}{\natexlab{a}}).

\bibitem[{\citenamefont{Fedichev et~al.}(1996)\citenamefont{Fedichev, Reynolds,
  and Shlyapnikov}}]{Fedichev1996tbr}
\bibinfo{author}{\bibfnamefont{P.~O.} \bibnamefont{Fedichev}},
  \bibinfo{author}{\bibfnamefont{M.~W.} \bibnamefont{Reynolds}},
  \bibnamefont{and} \bibinfo{author}{\bibfnamefont{G.~V.}
  \bibnamefont{Shlyapnikov}}, \bibinfo{journal}{Phys. Rev. Lett.}
  \textbf{\bibinfo{volume}{77}}, \bibinfo{pages}{2921} (\bibinfo{year}{1996}).

\bibitem[{\citenamefont{Weber et~al.}(2003{\natexlab{b}})\citenamefont{Weber,
  Herbig, Mark, Nagerl, and Grimm}}]{Weber2003tbr}
\bibinfo{author}{\bibfnamefont{T.}~\bibnamefont{Weber}} {\it et al.},
  \bibinfo{journal}{Phys. Rev. Lett.} \textbf{\bibinfo{volume}{91}},
  \bibinfo{eid}{123201} (\bibinfo{year}{2003}{\natexlab{b}}).

\bibitem[{\citenamefont{N\"{a}gerl et~al.}(2006)\citenamefont{N\"{a}gerl,
  Kraemer, Mark, Waldburger, Danzl, Engeser, Lange, Pilch, Jaakkola, Chin, Grimme}}]{Naegerl2006eef}
\bibinfo{author}{\bibfnamefont{H.-C.} \bibnamefont{N\"{a}gerl}} {\it et al.},
  \bibinfo{journal}{Atomic Physics 20 AIP Conf. Proc.}
  \textbf{\bibinfo{volume}{869}}, \bibinfo{pages}{269} (\bibinfo{year}{2006}),
  \bibinfo{note}{cond-mat/0611629}.

\bibitem[{\citenamefont{Westfall and Young}(1993)}]{Westfallbook}
\bibinfo{editor}{\bibfnamefont{P.~H.} \bibnamefont{Westfall}} \bibnamefont{and}
  \bibinfo{editor}{\bibfnamefont{S.~S.} \bibnamefont{Young}}, eds.,
  \emph{\bibinfo{title}{Resampling-Based Multiple Testing}}
  (\bibinfo{publisher}{John Wiley and Sons, New York}, \bibinfo{year}{1993}).

\bibitem[{\citenamefont{Mehta et~al.}(2009)\citenamefont{Mehta, Rittenhouse,
  von Stecher, D'Incao, and Greene}}]{Mehta}
\bibinfo{author}{\bibfnamefont{N.~P.} \bibnamefont{Mehta}},
  \bibinfo{author}{\bibfnamefont{S.~T.} \bibnamefont{Rittenhouse}},
  \bibinfo{author}{\bibfnamefont{J.}~\bibnamefont{von Stecher}},
  \bibinfo{author}{\bibfnamefont{J.~P.} \bibnamefont{D'Incao}},
  \bibnamefont{and} \bibinfo{author}{\bibfnamefont{C.~H.}
  \bibnamefont{Greene}}, \bibinfo{note} {\emph{A general theoretical description of $N$-body recombination}}, \bibinfo{journal}{unpublished}
  (\bibinfo{year}{2009}).

\bibitem[{\citenamefont{Zaccanti}(2009)}]{zaccanti2009}
\bibinfo{author}{\bibfnamefont{M.}~\bibnamefont{Zaccanti}},
  \bibinfo{journal}{private comunication}  (\bibinfo{year}{2009}).

\end{thebibliography}

\end{document}